\documentclass[aps,prl,twocolumn,showpacs,superscriptaddress]{revtex4}
\usepackage{graphicx}
\begin{document}

\title{Comment on ``Tracer Diffusion in a Dislocated Lamellar System''} 

\author{Doru Constantin}
\email{dcconsta@ens-lyon.fr}
\affiliation{Laboratoire de Physique de l'ENS de Lyon, 46 All\'ee d'Italie, 69364 Lyon Cedex 07, France}
\author{Robert Ho{\l}yst}
\email{holyst@ichf.edu.pl}
\affiliation{Institute of Physical Chemistry PAS, Dept. III, Kasprzaka 44/52, 01-224 Warsaw, Poland}



\maketitle

In a recent Letter, Gurarie and Lobkovsky \cite{gurarie02} affirm that the movement of tracer particles across the layers in a lamellar system with screw dislocations is super-diffusive (ballistic when dislocations of one sign are in excess and with normal displacement $\langle z^2(t) \rangle \propto t \log t$ when the average dislocation charge is zero). We argue that, when surface geometry is properly taken into account, the normal displacement is diffusive and the associated diffusion constant $D_{\bot}$ is always smaller than the in-plane diffusion constant $D$.

Particle diffusion on a curved surface extending indefinitely in all three directions in space is characterized by a displacement $\langle x^2 + y^2 + z^2 \rangle = D_{\text{eff}} t$, where the effective diffusion coefficient $D_{\text{eff}} \leq D$, with $D$ the local diffusion coefficient in the plane of the surface \cite{holyst}. Equality is achieved when the surface has zero mean curvature everywhere (minimal surface) \cite{anderson90}. Thus, the displacement along any particular direction (say $z$) has an upper bound
\begin{equation}
\langle z^2(t) \rangle \leq Dt
\label{eq:zsq}
\end{equation}
This can be shown by taking a cutoff at a finite distance from the core (as done in \cite{gurarie02}) and introducing reflecting boundary conditions on a helix at distance $a$ from the core. The surface is smooth and one can perfectly define the diffusion equation locally. As discussed in \cite{anderson90}, the diffusion tensor is diagonal in a local reference frame, of values $D$ in the plane of the surface and $0$ along the normal to the surface \footnote{Note the factor $3/2$ between our definition of $D$ and their definition of $D_0$.}. Thus, diffusion cannot be faster than $D$ in any particular space direction $\hat{z}$, as one can see by averaging $ \hat{z} \overline{\overline{D}} \hat{z}$ over an arbitrary surface. The limit is given by the {\em local} properties of the surface and does not depend on its detailed overall configuration. In reference \cite{gurarie02}, the authors artificially decouple the coordinates so that the particle diffuses in the $(x,y)$ plane, but the ``cost'' associated with displacement along $z$ is not taken into account, leading to the violation of (\ref{eq:zsq}).

\begin{figure}[h]
\includegraphics{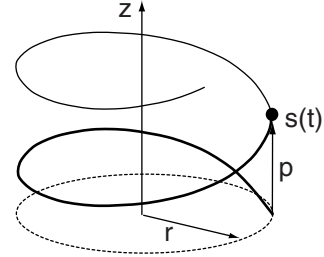}
\caption{\label{fig:helix}Helical trajectory of radius $r$ and pitch $p$ around a screw dislocation (the core is along the $z$ axis). The position of the particle is given by the curvilinear coordinate $s(t)$, with $s=0$ in the $z=0$ plane.}
\end{figure}

To illustrate these observations on a very simple case, let us consider the case of a tracer particle that is constrained to remain at a distance $r$ from the dislocation core (FIG.~\ref{fig:helix}) and consequently diffuses along a helix with pitch $p$ equal to the lamellar spacing.

Suppose the particle starts at $z=0$ at time $t=0$; the statistical distribution of its curvilinear coordinate along the helix $s(t)$ is then described by~: $\langle \Delta s(t) \rangle = 0$ and $\langle \Delta s^2(t) \rangle = 2Dt$. When it moves one layer up (to $z=p$), its coordinate is $s=\sqrt{(2 \pi r)^2+p^2}$. It is then readily shown that~:
\begin{equation}
D_{\bot} = D \frac{p^2}{(2 \pi r)^2+p^2}
\label{eq:D_bot}
\end{equation}
Clearly, the particle always exhibits normal diffusion along $z$, with a diffusion constant $D_{\bot} \leq D$, even when $r \rightarrow 0$ (incidentally, in this limit one obtains the ``pipe diffusion'' along the dislocation core). On the contrary, if one neglects the $p^2$ term in the denominator of equation (\ref{eq:D_bot}) (as in reference \cite{gurarie02}), the ratio $D_{\bot}/D$ artificially diverges with vanishing $r$. This example is very relevant to the discussion in \cite{gurarie02} because, as the authors acknowledge, trajectories tightly wound around the core dominate the statistics. In our opinion, this is the cause of the divergence in their equation (11), and not the fact that the trajectories enclose several dislocations, as they imply.


R. H. was supported by KBN grant 2P03B00923.

\end{document}